\documentstyle[natbib,epsfig]{jfm}    
\newcommand{\be}{\begin{equation}}
\newcommand{\ee}{\end{equation}}
\newcommand{\bR}{{\bf R}}
\newcommand{\br}{{\bf r}}

\title{Effective pseudo-potentials of hydrodynamic origin}

\author[Todd M. Squires]{T\ls O\ls D\ls D\ns M.\ns S\ls Q\ls U\ls I\ls R\ls E\ls S}

\affiliation{Department of Physics, Harvard University, Cambridge, MA 02138}

\pubyear{2001}
\volume{?}
\pagerange{?--?}
\date{7 February 2001}
\begin{document}
\maketitle
\begin{abstract}
It is shown that low Reynolds number fluid flows can cause suspended particles to respond 
as though they were in an equilibrium system with an effective potential.  
This general result follows naturally from the fact that different
methods of moving particles in viscous fluids give rise to very different
long-range flows.  Two examples are discussed:  electrophoretic `levitation' of 
a heavy charged sphere, for which a hydrodynamic `pseudo-potential' can be written in
closed form, and quasi-two dimensional crystals of like-charged colloidal spheres which form near charged walls, whose apparent attraction arises not from a force but from persistent 
fluid flows.  
\end{abstract}

\section{Introduction}

It is well-known that a particle which is moving under the influence of an external force such as gravity 
in a viscous fluid sets up a disturbance flow in the fluid which, in the low Reynolds number limit, 
decays with distance like $r^{-1}$.  The long-range character of these
force-driven flows manifests itself in many situations. 
One example involves collective diffusive behavior.  A pair of Brownian spheres has been
predicted \cite*[]{bat76} and measured \cite*[]{crocker:I} to diffuse in a correlated manner.  
On the basis of these long-range correlations, \cite{crocker:II} have developed a new technique for 
the microrheology of  complex materials.  Recently, \cite{dufresne} experimentally and
theoretically showed the diffusive behavior of a pair of Brownian
spheres to be strongly influenced even by distant solid boundaries.
Long-range flows also have a significant influence on the properties of sedimenting systems,
giving rise to numerous divergent integrals \cite[see e.g.][]{hinch2}.  For example, 
\cite{caflisch} have predicted velocity fluctuations in sedimenting particulate suspensions
to diverge with system size, and \cite{brenner} has drawn attention to the fact that sedimenting particles' 
long-range interaction with cell walls can screen the flows and cut off the divergent fluctuations.  
In addition, long-range hydrodynamic coupling has provided a plausible interpretation 
\cite[]{squires} for the experiment of \cite{larsen}, which measured an apparent attraction 
between like-charged colloidal spheres in the proximity of a similarly-charged wall.

It is also well-known that there are other ways to move particles 
which do not give rise to such long-range disturbance velocity fields.
One example is electrophoresis \cite*[e.g.][]{anderson}, wherein charged 
particles move under 
the influence of an applied electric field.  A charged particle in solution attracts
oppositely charged counter-ions from the solution, which form a screening `double-layer'
\cite[e.g.][chap. 4]{saville}.  The particle/double layer system is electrically 
neutral, so an applied 
electric field exerts no net force on the ensemble.  However, the charged particle and the 
counter-ions are driven in opposite directions, giving an apparent `slip velocity'
at the edge of the double layer, and a net motion to the particle.  Since velocity fields 
with $r^{-1}$ decay arise only 
from a net force on the fluid, electrophoretic (and all force-free) flow fields must 
decay at least as fast as $r^{-2}$.  

In this paper we will demonstrate that this difference between the long-range 
disturbance flows due to forced and force-free motions can be exploited to cause particles 
to behave as though they were subject to equilibrium forces derived from an
effective potential.  As an example, we will discuss the electrophoretic 
`levitation' of a heavy charged sphere off a wall, for which an effective 
pseudo-potential $\Phi_{{\mbox{ps}}}$ can be determined exactly.  We will then present a method for 
making like-charged colloidal spheres near a wall behave as though there 
was an apparent attraction between them, so that layered like-charged colloidal
crystals can be grown.  Throughout this work, we use the term `pseudo-potential'
to emphasize that the hydrodynamic component of $\Phi_{{\mbox{ps}}}$ arises not 
from a force, but from persistent low-Reynolds number flows which entrain each particle.

We shall first briefly review the hydrodynamics of forced- and force-free motion.
Colloidal particles ($\sim 1\mu m$) and their typical velocities ($\sim 1\mu m/s$) are 
small enough that inertial effects are entirely negligible.  In this limit, a rigid 
sphere of radius $a$ moves due to an external force 
${\bf F}$ at velocity ${\bf U_0} = b_0 {\bf F}$, where $b_0=(6 \pi \eta a)^{-1}$ is the Stokes mobility. 
This motion sets up a disturbance flow in the surrounding fluid
\be
{\bf u}({\bf R}) = \frac{1}{8 \pi \eta} \left(\frac{{\bf I}}{R} + 
\frac{{\bf R R}}{R^3} \right)\cdot {\bf F}-
\frac{a^2}{24 \pi \eta}\left(-\frac{{\bf I}}{R^3} + 
\frac{{3\bf R R}}{R^5} \right)\cdot {\bf F},
\ee
where $\eta$ is the fluid viscosity and ${\bf R} ={\bf r}-{\bf r}_0$ is the vector connecting the
observation point ${\bf r}$ and the center of the sphere ${\bf r}_0$.  This can be re-expressed
using fundamental singularities of the Stokes equation as
\be
{\bf u}({\bf r}) = \frac{1}{8 \pi \eta} \left\{
{\bf S}( {\bf R} )  - \frac{a^2}{3} 
\nabla \nabla \cdot \left(\frac{1}{R} \right)  
\right\}\cdot {\bf F},
\ee
The first term represents the flow due to a point force, known as a Stokeslet, which decays like $R^{-1}$.  
The second term is the Green's function for a point source dipole, also called a potential dipole, 
which decays like $R^{-3}$.  The flow far from a forced sphere is
dominated by the Stokeslet, which depends on neither the size nor the velocity of the sphere.

By contrast, a charged insulating sphere with a thin double-layer moving due to a 
uniform electric field ${\bf E_\infty}$ sets up a disturbance flow outside the double layer
\cite[e.g.][p. 256]{saville},
\be
{\bf u}_E({\bf R}) = \frac{\epsilon \zeta}{4 \pi \eta}\frac{a^3}{2}\left(-\frac{{\bf I}}{R^3} + 
\frac{3{\bf R R}}{R^5} \right)\cdot {\bf E_\infty},
\ee
where $\epsilon$ is the dielectric constant of the fluid and $\zeta$ is the potential difference
across the double-layer.  Here the sphere moves at velocity ${\bf V_0}=M_0 {\bf E_\infty}$,
where $M_0 = \epsilon \zeta/ 4 \pi \eta$ is the electrophoretic mobility.  This flow is
exclusively potential dipole flow and decays like $R^{-3}$.  

Walls affect these flows significantly, since the fluid velocity must vanish
identically on solid boundaries.  
Image singularities, analogous to image charges in 
electrostatics, can often be found which exactly cancel a flow set up on the wall.
In this paper, we shall make extensive use of the flow due to a Stokeslet (point force) 
$F {\bf \hat{z}}$ oriented perpendicular to a wall, which \cite{blake} found to be
\be
{\bf u}(\bR) = \frac{F}{8 \pi \eta} \left(
{\bf S}(\bR) - {\bf S}(\bR^i) 
+ 2 h^2\left. \left\{ \nabla_0 \nabla_0
\left( \frac{1}{r} \right) \right\} \right|_{{\bf R}^i}
 - 2h \left. \left\{  \frac{\partial {\bf S}(\br)}{\partial z_0} 
\right\} \right|_{{\bf R}^i}\right)\cdot {\bf \hat{z}}, 
\label{eq:images}
\ee
where $\bR^i=\bR+2 h {\bf \hat{z}}$ is the vector between the images and the observation point,
derivatives with respect to the source point ${\bf r}_0$ (denoted by $\nabla_0$) should be taken first, 
and then the position ${\bf R}^i$ inserted in place of ${\bf r}$. 
The flow due to the images decays like $R^{-1}$, whereas the image flow for a source dipole 
\cite[]{blacha}, which will not be required in this paper, decays like $R^{-3}$.  

The velocity of a sphere forced toward a wall has two contributions: it moves at ${\bf U_0}$ due to the 
applied force,
and it is advected with the flow set up by the image singularities.  Evaluating the image flow from 
(\ref{eq:images}) at $r_0$ results in an $O(a/h)$  wall correction to the (scalar) mobility 
for motion perpendicular to the wall \cite[e.g.][]{happel}:
\be
b_\perp (h) = b_0 \left(1-\frac{9a}{8h} +O\left[\frac{a^3}{h^3}\right]\right). 
\label{eq:hydromobility}
\ee
The correction to the electrophoretic mobility can be derived in similar fashion, but is
slightly more complicated because the wall gives both hydrodynamic and electrostatic contributions.
Since the image flow (and electrostatic image potential) for electrophoretic motion decays 
like $R^{-3}$, the wall correction to the electrophoretic mobility is 
$O(a^3/h^3)$, and has been calculated using the method of reflections by \cite{keh} to give
\be
M_\perp(h) = M_0 \left( 1-\frac{5a^3}{8h^3} +O\left[\frac{a^5}{h^5}\right] \right).
\label{eq:electromob}
\ee
Throughout this paper, we will treat $a/h$ and $a/r$ as small, and so neglect this $O(a^3/h^3)$ 
variation in $M_\perp$.

\section{Electrophoretic `levitation'}

These two motions and their respective flows can be superposed, since Stokes' 
flow is linear.  A charged sphere which sediments due to a force $F_{\rm{g}}$ towards
a planar electrode and which is electrophoretically driven upwards due to a uniform
electric field ${\bf E_\infty}$ will have a
velocity $v$ given by
\be
v = -b_\perp (h) F_{\rm{g}} + M_\perp(h) E_\infty,
\ee 
with $b_\perp(h)$ and $M_\perp(h)$ given by (\ref{eq:hydromobility}) and (\ref{eq:electromob}) respectively.
If the bulk electrophoretic velocity is less than the bulk sedimentation velocity,
so that $\psi=M_0 E_\infty/b_0 F_{\rm{g}}<1$, there will be a unique height
\be
\frac{h_0}{a} = \frac{9}{8( 1-\psi)} +O(a^2)
\ee
where the two velocities exactly balance and a non-Brownian sphere will come to 
rest.  A sphere started above $h_0$ falls faster than electrophoresis
drives it up, and it moves down towards the plane.  Similarly, a sphere below 
$h_0$ falls too slowly, and electrophoresis drives it up.  Thus $h_0$ is a stable 
steady location.  On the other hand, $h_0$ would be an unstable steady point for a sphere
falling away from a wall.

A sphere undergoing Brownian motion will have a probability distribution $P(h)$
of being located at $h$ which is given by the steady Fokker-Planck equation \cite[see e.g.][]{batch2}
\be
\frac{\partial P}{\partial t} =0=-\nabla \cdot {\bf J} \equiv - \nabla \cdot \left[{\bf v}(h) P - k_B T 
{\bf b}(h) 
\cdot \nabla P \right],
\label{eq:fokker}
\ee
where $k_{\rm{B}} T {\bf b}$ is a tensor diffusivity.  This equation has solution
$P(h) = P_0 \exp(-\Phi_{{\mbox{ps}}}/k_B T)$,
where $\Phi_{{\mbox{ps}}}$ is a non-equilibrium `pseudo-potential'
\be
\Phi_{{\mbox{ps}}}(h) = F_{\rm{g}} h - \int^h \frac{M_\perp(h') E_\infty}{b_\perp(h')} dh'.
\label{eq:pseudo}
\ee
Inserting the values of $b_\perp$ and $M_\perp$ from (\ref{eq:hydromobility}) and (\ref{eq:electromob})
and keeping terms to $O(a/h)$, we integrate (\ref{eq:pseudo}) to obtain 
\be
\Phi_{{\mbox{ps}}}(h)  \approx F_{\rm{g}} h - \frac{M_0 E_\infty}{b_0} \left[h + \frac{9 a}{8}
\ln \left( \frac{h}{a} - \frac{9}{8} \right) \right],
\ee
which can be expanded about its minimum at $h_0$ to give an approximately harmonic well 
\be
\frac{\Phi_{{\mbox{ps}}}}{k_B T} \sim \frac{8}{9}\frac{F_{\rm{g}} a}{k_B T}\left[ \frac{(1-
\psi)^2}{\psi} \right]
\frac{(h-h_0)^2}{2a^2}+\frac{\Phi_{{\mbox{ps}}}(h_0)}{k_B T} .
\ee
If one viewed only the motion of a single sphere in this configuration, one would not be able to distinguish between 
this {\sl non-equilibrium} pseudo-potential and a true thermodynamic potential.  However, the
physical distinction is crucial:  the {\sl only} force on the sphere/double layer system
is $F_{\rm{g}}$, directed toward the wall.

\section{Multi-particle pseudo-potential interactions}

Even though the electrophoretically `levitated' particle is stationary, it drives 
a long-range flow.  Far from the sphere, the flow is always dominated by the 
force-driven motion, since it decays the most slowly.  Streamlines for the flow around
a hovering sphere, described in section 2, are shown in figure 1.

\begin{figure}[htbp]
\centerline{\epsfysize=3.5in\epsfbox{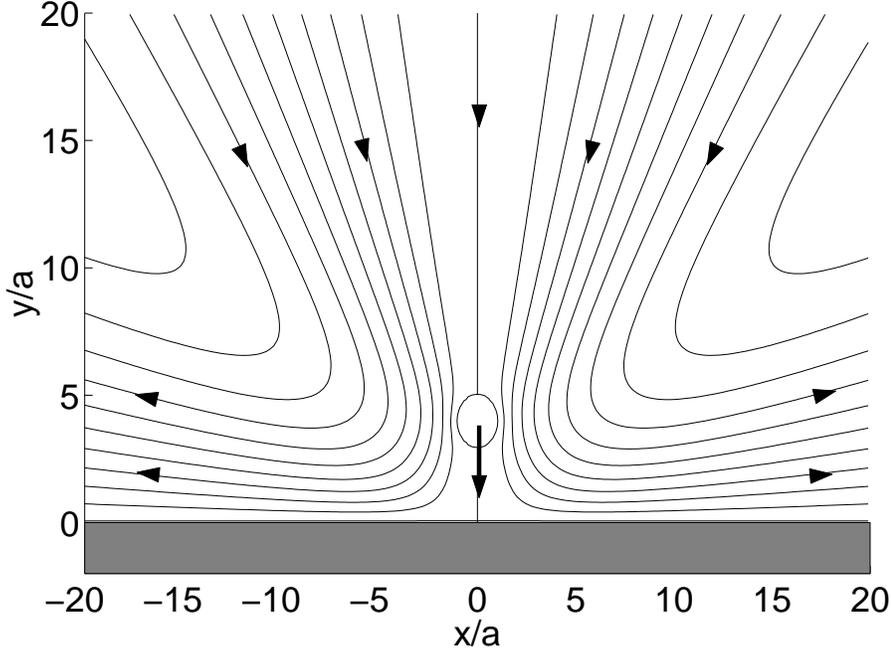}}

\caption{Streamlines for a heavy charged sphere electrophoretically levitated
at $h_0 = 4 a$.  Flows move down and radially out along the streamlines.  The closed 
streamline shows the slip velocity on the sphere surface.  The 
flow decays slowly and asymptotically approaches the flow due to a falling sphere 
near a wall, even though the sphere is immobile.  Gravity exerts a force directed 
toward the wall, and given a sphere with positive $\zeta$ potential, the electric field 
is directed away from the wall.
}
\label{streamlines}
\end{figure}

A second identical sphere, placed at the same height and at some distance $r$ away from 
the original sphere, will both set up a long-range flow as well as be 
entrained by the flow set up by the first sphere.  Forcing two spheres perpendicular to the wall 
induces a relative velocity between them \cite[]{squires}, in the plane of the wall, as
is evident from the streamlines in figure 1.  This relative velocity of sphere 1 arises from 
its entrainment in the flow set up by the images of sphere 2, and is found to $O(a/h)$ and $O(a/r)$ 
from (\ref{eq:images}) to be
\be
{\dot x_1}=-{\dot x_2}=- \frac{3}{2\pi \eta} \frac{r h^3}{(r^2+4 h^2)^{5/2}} F_g,
\label{eq:rdot}
\ee
where $x$ is the coordinate parallel to the wall, along the line joining the spheres, as in figure 2.
Two spheres forced away from a wall move together as they leave the wall, whereas two spheres 
falling towards a wall move apart.  In this case, we observe that the relative motion between 
two non-Brownian hovering 
spheres resembles the motion due to an effective repulsive force, even though it 
does not arise from a force.  
Example trajectories for two such spheres are shown by the solid curves in figure 2, with
arrows showing the collapse onto these trajectories from various initial configurations.  

\begin{figure}[htbp]
\centerline{\epsfysize=3.5in\epsfbox{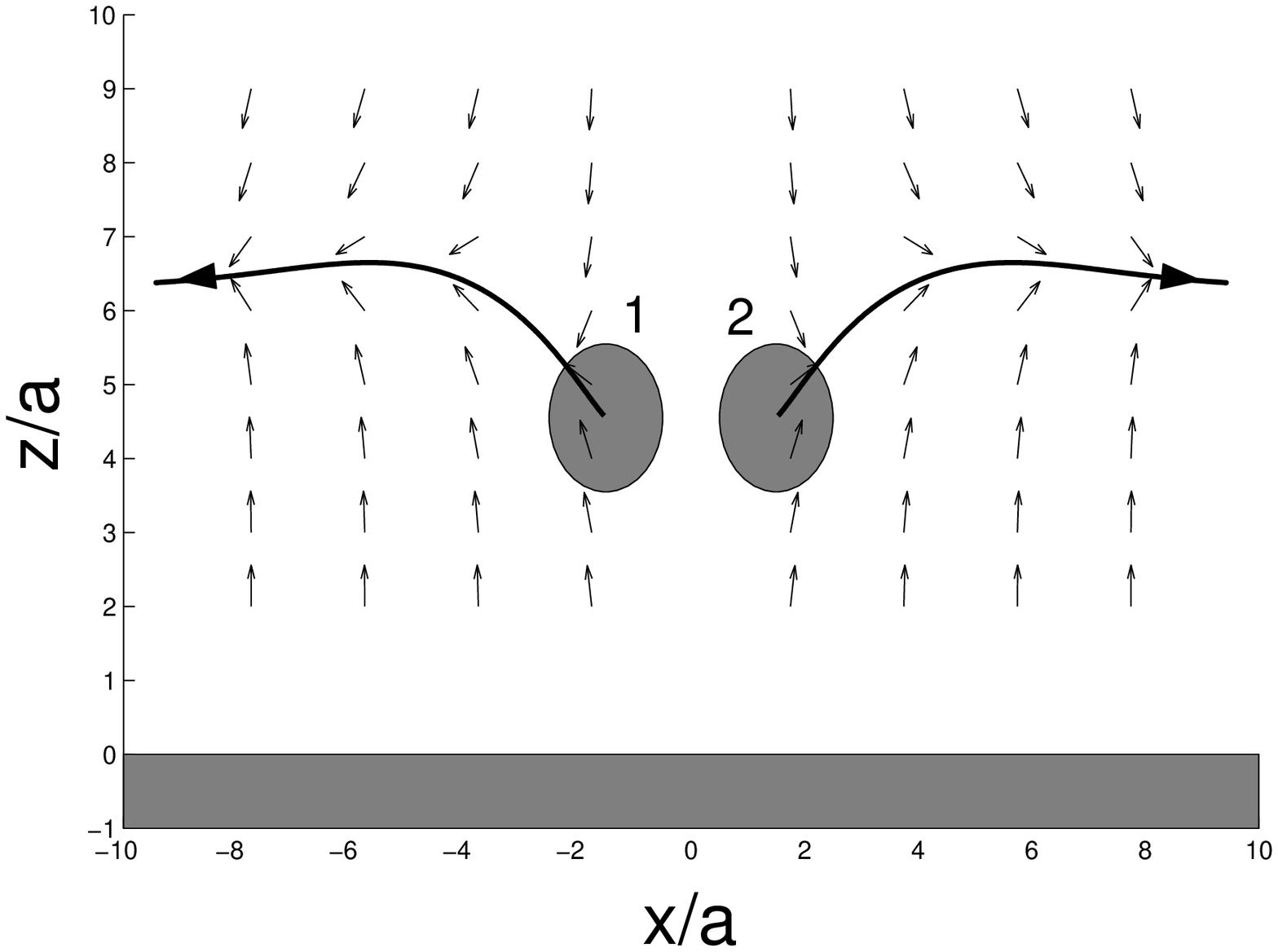}}
\caption{Two heavy spheres electrophoretically driven 
upwards approach a stable trajectory, on which the spheres move apart.
This mimics a repulsion between the spheres, and asymptotes to two 
spheres levitating independently.  In this figure, we have chosen $E$ such
that an independent sphere would hover at $h=6a$.}
\end{figure}

\section{Like-charge colloidal crystal formation}

Finally, we explore the possibility of setting up an attractive interaction between
the spheres, which can lead to crystal formation.  Since forcing two spheres toward a wall
led to an effective repulsion, forcing them off of the wall will give an effective attraction.
One way to do this would be to electrophoretically drive neutrally buoyant charged spheres into a 
similarly-charged wall.  This would require the wall to possess a static surface charge density and to simultaneously passes a current.

As we have seen, the far-field hydrodynamic flows are dominated by the net force
on the body and its charge cloud.  In this case, the net force $F_{\rm{w}}$ on each is given by 
its electrostatic repulsion from the static charge on the wall, found by integrating an electrostatic
stress tensor over the surface of the sphere \cite[see, e.g.][]{saville}.  Under the linear superposition
approximation, the repulsive force the wall exerts on the sphere is found to be
\begin{equation}
\frac{F_{\rm{w}}}{k_{\rm{B}}T}=4 \pi Z \sigma_{\rm{g}}\lambda _{\rm{B}}\frac{e^{\kappa a}}{
1+\kappa a}e ^{-\kappa h},
\label{wallforce}
\end{equation}
where $Z$ is the effective charge on the sphere, $\sigma_{\rm{g}}$ is the charge density on the wall, $\kappa^{-1}$ is
the Debye screening length and $\lambda_{\rm{B}}$ is the Bjerrum length($\approx .7$ nm in water at room temperature).
This force sets up a flow around each sphere like in figure 1, with flow arrows reversed.  The 
lateral component of the flow now draws the spheres together.  

Two non-Brownian spheres 
separated by distance $r$ are immobilized at the height $h_0$ where 
${\bf M \cdot E}_\infty = {\bf b}(h_0,r) {\bf \cdot F}_{\rm{w}}(h_0)$.
The equivalent force on each sphere which would be necessary to drive this motion in a quiescent fluid is
\be
F_{x_1} \approx  \frac{\dot{x_1}}{b_{x_1x_1} - b_{x_1x_2}},
\ee
where $b_{x_1x_2} $ is the coefficient of ${\bf b}$ which gives the mobility sphere 1
in the $\hat{{\bf x}}$ direction due to a force on sphere 1 in the same direction, and so on.
Using (\ref{eq:rdot}), and neglecting the 
weak dependence of $h_0$ upon $r$, this can be integrated to yield a pseudo-potential
\be
\Phi_{\mbox{ps}}(r,h_0)\approx U_{\rm{p}}(r)-
\frac{3 h_0^3 F_{\rm{w}}(h_0)}{2 \pi \eta}\int^r \frac{r(4h_0^2 + r^2)^{-5/2}}
{b_{x_1x_1}(h_0) - b_{x_1x_2}(r,h_0)} dr,
\label{eq:pseudoint}
\ee
where $U_{\rm{p}}$ is the true pair potential between the spheres.  Although
$b_{x_2x_1}$ is $O(a)$, in practice it is small compared to $b_{x_1x_1},$ 
and its neglect allows the direct integration of (\ref{eq:pseudoint})
to give
\be
\Phi_{\mbox{ps}}(r,h_0)\approx U_{\rm{p}}(r)-\frac{F_{\rm{w}}(h_0)}{1-\frac{9 
a}{16 h_0}}\frac{3 h_0^3 a}{(4h_0^2 +
r^2)^{3/2}}.
\label{eq:approx}
\ee
The true pair potential between charged spheres, $U_{\rm{p}}$, is given in linear 
superposition approximation by \cite[e.g.][]{larsen},
\begin{equation}
\frac{U_{\rm{p}}}{k_{B}T}=Z^{2}\lambda _{B}\left( \frac{e^{\kappa a}}{1+\kappa
a}\right) ^{2}\frac{e^{-\kappa r}}{r}.
\label{dlvogrier}
\end{equation}

Equation (\ref{eq:approx}) is identical to that derived by \cite{squires} in their analysis of the 
experiment of \cite{larsen} which measured the effective interaction between like-charged colloidal spheres
near a single charged wall.  Although the two formulae were derived in the same fashion, they
have very different physical consequences.  In that experiment, 
a pair of spheres was repeatedly released from the same height $h_0$ and then re-trapped; therefore,
a net force systematically drove the spheres away from the wall throughout the experiment.  
Consequently, there was a systematic relative motion of the spheres towards each other which had been interpreted 
to arise from an attractive force.  In an equilbrium system, where the spheres fluctuate into the wall as often
as away, one would expect no systematic relative motion, and this hydrodynamic effect 
would be of no physical consequence.

Similarly, the pseudo-potential of equation (\ref{eq:approx}) does not describe an attractive force 
between two like-charged 
colloids.  However, in the present situation, similarly-charged spheres actually do respond as though 
there was an attraction.  The effect of electrophoresis is to `pin' the spheres at some height without
applying a force, so that the long-range flow is dominated by the forced motion and entrains neighboring
spheres accordingly.
A collection of spheres can thus aggregate to form clusters--a behavior which might otherwise seem 
to arise from an attractive force.

Spheres which are not neutrally buoyant have a similar pseudo-potential,
with $F_{\rm{w}} - F_{\rm{g}}$ in place of $F_{\rm{w}}$ in (\ref{eq:approx}).  Note that both the 
depth and the range of this potential can be tuned independently--the range by adjusting 
the electric field $E_\infty$ to give different heights $h_0$ and the overall magnitude 
of the attraction by adjusting the relative density of the particle to affect $F_{\rm{g}}$.  
Furthermore, one could study the freezing and melting of the crystals we'll describe next by
simply adjusting $E_\infty$ during an experiment.

Brownian motion considerably complicates the above picture.  Again, the 
probability distribution for the spheres' positions obey the steady 
Fokker-Planck equation (\ref{eq:fokker}).  
An interesting fact is that there is a persistent non-zero probability current 
${\bf J}$ in the steady state. The probability current can only be zero when
\be
 - \nabla \Phi_{{\mbox{ps}}}= {\bf b}^{-1}\cdot {\bf u}
\label{eq:flux}
\ee
has a solution.  In equilibrium systems, ${\bf u} = {\bf b}\cdot {\bf F}$
so that ${\bf F}= -\nabla \Phi_{{\mbox{ps}}}$ as expected.  Here 
${\bf u} = {\bf b}\cdot {\bf F} + {\bf u}_E$, and (\ref{eq:flux})
will not in general have a solution.  Physically, this is because the equivalent 
force required to give velocity ${\bf u}_E$ would be non-conservative.  Vertical fluctuations
take the pair into regions of stronger or weaker attraction, leading to persistent
closed orbits of ${\bf J}$. 
When these vertical fluctuations are relatively small, ${\bf J}$ is small and the 
effective pseudo-potential is well approximated by (\ref{eq:approx}).  A 
comparison betweeen (\ref{eq:approx}) and simulated results is shown in figure 3.  The lower two
pseudo-potential wells (a) and (b) are deep enough that bound states form. The simulated pseudo-potentials
$\Phi_{{\rm ps}}/k_B T$ are thus simply $-\log \ls g(r)$, where $g(r)$ is the the pair correlation function
obtained from simulations of a pair of spheres.  
We used the method described by \cite{crocker:III} to simulate the shallowest potential well (c),
since it is not deep enough for bound states. 

We have run simulations which account for Brownian motion to demonstrate that 
spheres can indeed self-assemble into a quasi-2D crystal.  We use the 
dynamical equations 
\be
{\bf v} = {\bf b} \cdot {\bf F} + {\bf M}\cdot {\bf E}
\ee
where ${\bf v}, {\bf b}$ and so on are multiparticle quantities.  We retain terms 
up to $O(a)$ in ${\bf b}$ and ${\bf M}$ and use a box of side $100a$ 
with periodic boundaries.
Brownian motion is accounted for in standard fashion \cite[]{ermak,hinch}, wherein ${\bf D} 
= k_B T {\bf b}$ is a position-dependent tensor diffusivity.  We use as input parameters the
values $Z=7300$, $\kappa^{-1}=
.275 \mu m$ and $2a=.652\mu m$ as in the experiments of \cite{larsen}
and the value $\sigma_{\rm{g}} = .4\sigma_s$ for the (unmeasured) charge density on the wall
which \cite{squires} found as the best fit to the experimental data. 
We choose these values because 1) the effects we are discussing are most pronounced with 
long screening lengths and high surface charge densities, so that the spheres tend to `hover'
at an appreciable distance from the wall, and 2) these effects may bear some 
relation to these particular experiments.  We started nineteen spheres in a hexagonal crystal, then
simulated their dynamics until it was clear that the qualitative features would not change.  This 
was typically approximately a minute or less of real time.
Figure 3 shows the resulting suspensions at the end of each run for three different
field strengths.  The strongest field, for which $h_0=2\mu m$, yields a hexagonal crystal, whereas
a weaker field, for which $h_0 = 2.2 \mu m$, is strong enough to yield a dense phase, but not
strong enough to retain orientational order.  We loosely term this an istropic `liquid' phase. 
Finally, for an even weaker field, for which $h_0 = 2.5 \mu m$, the attraction is sufficiently weak
that the spheres spread out to fill the box as an isotropic `gas'.
\begin{figure}[htbp]

\centerline{\epsfysize=3.5in\epsfbox{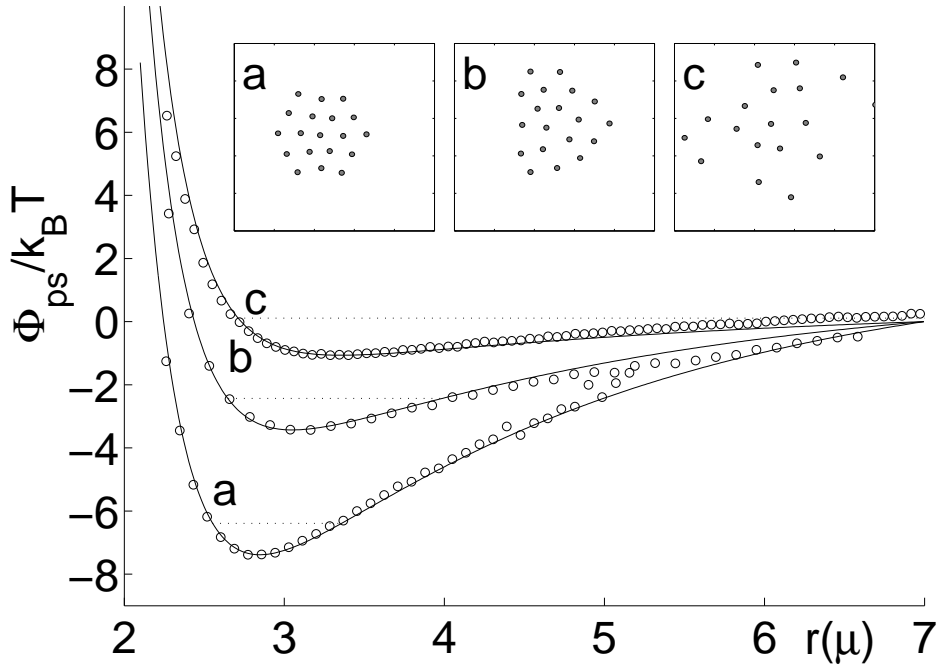}}

\caption{Radial pseudo-potentials and representative `phases' of neutrally buoyant 
charged spheres electrophoretically driven into the double-layer surrounding a 
charged wall.  a) A strong field ($h_0 = 2\mu m$) yields a deep pseudo-potential well
and tight pairwise confinement, giving a regular triangular lattice.  b) 
A moderate field ($h_0 = 2.2\mu m$) yields a pseudo-potential well deep enough to yield
a dense phase, but not strong enough to retain orientational order. c) A weak
field ($h_0 = 2.5\mu m$) yields an isotropic `gas'.  Dotted lines represent 1 $k_B T$ of 
energy and give typical excursions about the equilibrium point.  Lines give the 
approximate analytic form (\ref{eq:approx}), and dots give the simulated 
 pseudo-potentials.
}
\end{figure}

\section{Discussion and conclusions}

While we have specifically investigated the combination of electrophoretically-driven 
motion and body-force
motion, the present results have a wider validity.  Our results will hold whenever motion 
using body forces such as gravity, electrostatic repulsion or attraction, magnetic forces, 
etc. are combined with force-free
motion.  There are many ways to achieve the latter.  Since any self-propelled motion 
can not result in a net force or torque on the system, all swimmers generate flows decaying
faster than $r^{-1}$ \cite[]{lighthill}.  There are also other types of ``phoretic"  motion, which involve
particle motion due to fields which interact with the particle's 
surface \cite[e.g.][]{anderson}, such as 
thermophoresis (motion due to thermal gradients) and
diffusophoresis (motion due to solute concentration gradients).  Similarly, the 
thermocapillary motion of bubbles \cite[]{young}, driven by 
thermal surface tension gradients, is force-free.

Having shown that like-charged colloidal crystals can self-assemble into 
ordered crystals due to long-range 
persistent non-equilibrium hydrodynamic flows, it is natural to ask whether 
there could be any connection between
these crystals and those observed by \cite{larsen}.  These 
metastable crystallites formed near the charged glass walls of the 
experimental cells, and persisted for times orders of 
magnitude longer than purely repulsive Brownian spheres
should. This response can certainly be understood
within the present picture, and would
require a force-free motion to drive the spheres into the charge cloud of the 
walls.  A spurious current in the cell could do
this; however, it is unlikely that this explains Larsen and 
Grier's crystals, for two reasons:  1) there
are no electrodes on the top and bottom walls of their cells, and 2) the 
crystals were observed to form on both the top
and bottom walls, whereas a net current in the cell would likely drive the 
spheres to one wall or the other.  Another possible current source, suggested by D. S.
Fisher, is the dissociation of ionic groups on the glass surface.  Other
possible mechanisms for force-free motion include thermal or electrolyte 
concentration gradients; however, it is not clear which if any of these
mechanisms are present or responsible for the crystals.  It is generally 
believed that a novel long-ranged attraction between confined like-charged colloids, 
whose origin remains controversial, gives rise to this behavior \cite[e.g.][]{hansen}.
The present work provides a non-equilibrium mechanism whose origins are 
well-known and understood which would mimic an attractive interaction and could drive
like-charge colloidal crystallization, so its potential role should be investigated further.

Two-dimensional colloidal crystals have been formed on electrodes via 
electrophoretic deposition \cite[]{bohmer,trau}, 
and \cite{solomentsev} has proposed a electrohydrodynamic model, which bears some
similarity to 
the present work, for its explanation.  The crucial 
difference between the our work and that of \cite{solomentsev} and is that in the 
latter, the particles are first deposited on the electrode, and then electro-osmotic 
flows drive adjacent particles towards one another.  These electro-osmotic flows are much
shorter-ranged than the body-force driven flows described here, and are highly 
screened by the close proximity of the wall.  

In conclusion, we have used the difference in flow fields 
set up by force-driven motion and force-free motion to show that 
persistent long-range
viscous flowfields can be set up, resulting in particle motion which 
mimics an effective potential.  This
appears surprising at first, since the system is out of equilibrium, and yet
behaves as though it is an equilibrium
system with an effective potential which is hydrodynamically driven.  
Given the general nature of this effect, these persistent flowfields 
are likely to find various applications in self-assembly and 
microfluidic contexts.

\begin{acknowledgments}
This work was supported under the NSF Division of Mathematical Sciences grant DMS-9733030.
\end{acknowledgments}

\bibliographystyle{jfm}

\end{document}